# Hybrid Model For Intrusion Detection Systems

*Baha Rababah*
*Department of Computer Science*
*University of Manitoba*
*Winnipeg, Canada*
baha@cs.umanitoba.ca*(7863789)*

*Srija Srivastava*
*Department of Computer Science*
*University of Manitoba*
*Winnipeg, Canada*
srivast1@myumanitoba.ca*(7860562)*

*Abstract*— With the increasing number of new attacks on ever-growing network traffic, it is becoming challenging to alert immediately any malicious activities to avoid loss of sensitive data and money. Thus, making intrusion detection as one of the major areas of concern in network security. Anomaly based network intrusion detection technique is one of the most commonly used technique. Depending upon the dataset used to test those techniques, the accuracy varies. Most of the times this dataset does not represent the real network traffic. Considering this, this project involves analysis of different machine learning algorithms used in intrusion detection systems, when tested upon two datasets which are similar to current real-world network traffic (CICIDS2017) and an improvement of KDD'99 (NSL-KDD). After the analysis of different intrusion detection systems on both the datasets, this project aimed to develop a new hybrid model for intrusion detection systems. This new hybrid approach combines decision tree and random forest algorithms using stacking scheme to achieve an accuracy of 85.2% and precision of 86.2% for NSL-KDD dataset, and achieve an accuracy of 98% and precision of 98% for CICIDS2017 dataset.

## I. INTRODUCTION

Due to the rapid evolution of sophisticated attacks and zero-day vulnerabilities on computer networks (referred as intrusion), the detection of these intrusions has become an area of high priority and concern. Intrusion detection system helps to monitor the computer network and identify the attacks, unauthorized activities or any malicious activities. Thus, strengthening the security of computer systems and networks. The basic architecture of an intrusion detection system can be illustrated as Fig.1.

Intrusion detection systems (IDSs) can be broadly classified into two categories: signature-based and anomaly-based IDSs. Here, signature-based IDSs, also referred as misuse-based, look for a defined pattern (signature of an attack) to detect the known attacks. These signatures of known attacks are maintained in a database, which requires frequent updates to accommodate new types of attacks. This scheme is beneficial for the well-known attacks. On the other hand, anomaly-based IDSs maintain the normal behavior of network and system, and identify an attack as a deviation from normal behavior. Because of this feature, anomaly-based IDSs have the capability to identify zero-day (novel) attacks.

Focusing on anomaly-based IDSs, according to Sharafaldin et al. [1], most of the proposed intrusion detection approaches are tested using datasets (representing network traffic) that are out of date, such as KDD99. In other words, those datasets have several weaknesses such as unreliability, redundancy, and lack of traffic/attacks diversity and current real-world data. Also, Ahmed et al. [2] described that the behavior of normal network traffic is changing continuously. Therefore, to overcome this challenge the IDSs should also be tested upon the current network to make the detection system more robust.

Keeping the aforementioned discussion as the motivation, this project aims to analyze the performance of some of the popular machine learning algorithms that are used in anomaly-based intrusion detection systems when tested upon the latest datasets. This analysis uses two recently published datasets: CICIDS2017 and NSL-Kdd. As per the analysis result, we proposed a hybrid model for anomaly-based IDSs, which combines two machine learning algorithms decision tree and random forest using stacking scheme. Eventually achieving better and higher accuracy of intrusion detection.

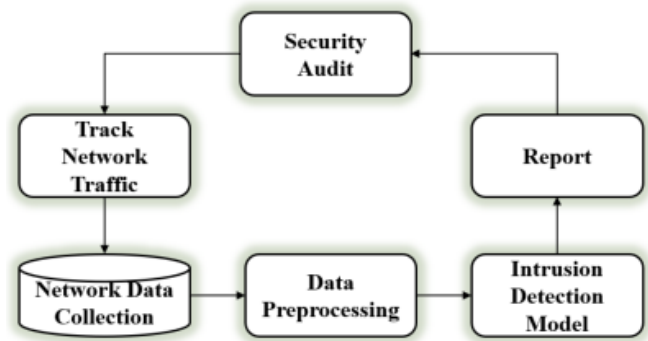

**Fig.1.** Intrusion Detection System Architecture

The rest of the paper is organized as follows: Section II focuses on some of the major related works in the area of intrusion detection, followed by Section III, which describes the tools and methodology used for different steps of IDSs' architecture (Fig.1) to facilitate the analysis. Approach for the new hybrid IDSs' model is specified in Section IV. Section V discusses the result of analysis for the machine learning algorithms used in IDSs and the performance of new model. The conclusion is presented in Section VI.

## II. RELATED WORK

The first model which was capable of detecting real-time intrusion was built in 1987 by Dorothy E. Denning [3],

where the author monitored the system's audit logs to identify the abnormal behavior. Thus, leading to further research in this area. After some exploration in this field, it was found in 2000, in a study [4] that the necessity of datasets (representing actual network systems) is gaining priority over DARPA datasets to test the intrusion detection systems being developed.

Later in 2002, Liao et. al. [5] demonstrated that k-Nearest Neighbor classifier (a machine learning algorithm) can effectively identify an attack, achieving low false positive rate. This system was tested upon 1998 DARPA dataset. Following this, in 2003 Mahoney et. al. [6] investigated and found that DARPA/MIT Lincoln laboratory evaluation dataset leads to an overoptimistic detection of network anomaly. The authors also suggested that this can be mitigated by mixing the real traffic with the simulated dataset.

Later, Zhang et. al. [7] proposed a framework which used random forest for misuse-based, anomaly-based and hybrid IDSs. Thus, achieving improved overall performance of intrusion detection, considering KDD'99 dataset for testing. Through all these years plethora of machine learning techniques have evolved resulting in better accuracy in detecting intrusions with lower false positive. An example of such evolution can be the hybrid technique proposed by Ravale et. al. [8] which combines K-means clustering and support vector machine (SVM) radial basis function (RBF) kernel. Also, Aljawarneh et.al. [9] in 2018, developed a hybrid intrusion detection model that combined J48, Meta Bagging, Random Tree, Decision Tree, AdaBoostM1, Decision Stump and Naïve Bayes, using vote scheme. Thus, achieving high detection accuracy of 99.81%.

Alongside these developments, several comparisons of performances of these intrusion detection techniques have been carried out. Belavagi et. al. [10] conducted tests for Logistic Regression, Gaussian Naïve Bayes, Support Vector Machine and Random Forest algorithms with NSL-KDD dataset. The author observed that Random Forest Classifier performs better than the other three algorithms (Table I).

TABLE I.

| Algorithms | Accuracy | Recall | Precision |
|---|---|---|---|
| Logistic Regression | 0.84 | 0.85 | 0.83 |
| Gaussian Naïve Bayes | 0.79 | 0.81 | 0.79 |
| Support Vector Machine | 0.75 | 0.79 | 0.76 |
| Random Forest | 0.99 | 0.99 | 0.99 |

Furthermore, in 2017 Almseidin et.al. [11] performed some experiments to evaluate the performances of machine learning algorithms namely, J48, Random Forest, Random Tree, Naïve Bayes, Decision Table, Bayes Network and MLP. When tested upon KDD dataset it was found that decision table had the lowest false negative value (0.002) but in terms of accuracy random forest outperforms (Table II). Similarly, a year later Zaman et. al. [12] performed tests to compare k-Means, k-Nearest Neighbors, Fuzzy C-Means, Support Vector Machine, Naïve Bayes, Radial Basis Function and Ensemble method combining all these six algorithms, for precision, accuracy and recall. These algorithms were tested upon Kyoto 2006+ dataset and it was found that Radial Basis Function was better compared to other algorithms (Table III).

TABLE II.

| Algorithms | Accuracy | Precision |
|---|---|---|
| J48 | 0.931 | 0.989 |
| Random Forest | 0.937 | 0.991 |
| Random Tree | 0.905 | 0.992 |
| Naïve Bayes | 0.912 | 0.988 |
| Decision Table | 0.924 | 0.944 |
| Bayes Network | 0.907 | 0.992 |
| MLP | 0.919 | 0.978 |

TABLE III.

| Algorithms | Accuracy | Recall | Precision |
|---|---|---|---|
| Naïve Bayes | 0.967 | 0.916 | 0.916 |
| K Means | 0.836 | 0.25 | 0.75 |
| Fuzzy C-Means | 0.836 | 0.25 | 0.75 |
| Support Vector Machine | 0.942 | 0.833 | 0.869 |
| Radial Basis Function | 0.975 | 0.958 | 0.92 |
| K-Nearest Neighbours | 0.975 | 0.916 | 0.956 |
| Ensemble | 0.967 | 0.958 | 0.884 |

III. METHODOLOGY

Considering the basic architecture of intrusion detection systems (Fig.1), prior to start with the analysis of machine learning algorithms used in IDSs, the following steps must be performed to get the reliable results:

- Selection of appropriate dataset which will be used for training and testing of intrusion detection algorithm. Since, our project involves analysis of algorithms on two recent close to real-world datasets, therefore we are using CICIDS2017 and NSL-KDD datasets which are explained in Section III.A.
- Perform data preprocessing such as attribute selection, to take into account only the useful information from the raw data along with many other advantages, which is explained in detail in Section III.B.
- After the preprocessed data are obtained, train and test all the algorithms to be evaluated using this data. The algorithms analyzed in this paper are explained in Section III.C.
- The results of experiments conducted are analyzed based on the following metrics (Section III.D): true positive, false positive, precision, recall and F-measure.

The above-mentioned steps are performed on Weka [13], an open source data mining software/tool. It is a powerful tool that facilitates processing of data, performing experiments for learning schemes, and visualizing data and results.

*A. Datasets*

As mentioned earlier this project considers two datasets for evaluation purpose, CICIDS2017 and NSL-KDD.

*1) CICIDS2017:* CICIDS2017 is a real-world network traffic IDS dataset which is publicly available *[14]* for use. This dataset contains benign along with the most up-to-date seven common attacks, which resembles the true real-world data *[1]*. The attacks included in this dataset are Brute Force attack, DoS, Web attack, Infiltration, Botnet attack, DDoS and PortScan attack.

CICIDS2017 contains more than 2 million records and 78 attributes. Out of these 2 million records, for this project we selected 88271 records as training data and 66203 records as testing data. This subset of CICIDS2017 dataset is selected randomly.

*2) NSL-KDD:* NSL-KDD is a publicly available *[15]* dataset that has resolved some of the inherent problems of the KDD'99 dataset *[16]*. KDD'99 had 78% of redundant records and 75% of duplicate records in the train and test data *[17]*, which was then fixed in NSL-KDD dataset. Thus, resulting in a non-biased result for intrusion detection. According to McHugh *[4]*, this dataset still lacks public network data.

NSL-KDD dataset have 42 attributes and we selected 125973 and 22544 records for training and testing purpose respectively. This dataset contains Denial of service (DoS), Remote to local (R2L), User to root (U2R) and Probe attacks.

*B. Attribute Selection*

Attribute selection is a process of selecting a subset of relevant features (variables/attributes) for use in model construction. Attribute selection process increases the utility of data by selecting only the useful features/attributes. Thus, by selecting only relevant features, the size of dataset reduces which facilitates better real-time training and testing time, and it also prevent biased predictions. All this eventually leads to the better accuracy of prediction result.

In this project we have used Information Gain Attribute Evaluation method to evaluate each attribute by measuring the information gain with respect to the class. Information Gain Attribute Evaluation method is a single-attribute evaluator which is used with Ranker search method, that ranks all the attributes according to their information gain. For attributes whose information gain value is below a certain threshold value specified, are removed from the dataset. In this project we have considered threshold value as 0.4. As a result of it, only 11 and 10 attributes are kept in NSL-KDD dataset (Fig. 3) and CICIDS2017 (Fig. 2) dataset respectively.

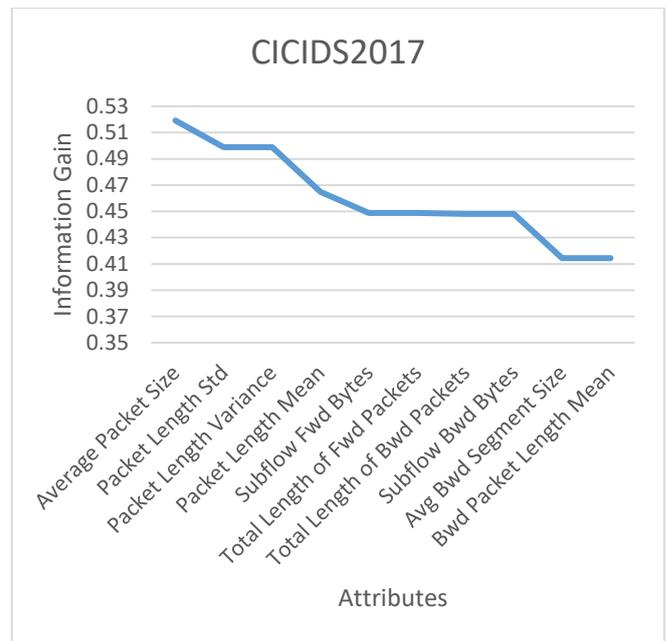

**Fig.2.** Information Gain Values of Attributes for CICIDS2017 dataset

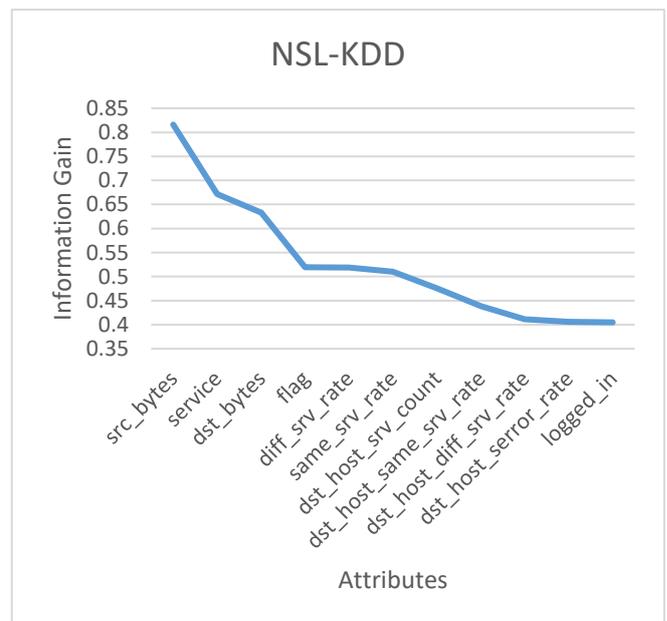

**Fig.3.** Information Gain Values of Attributes for NSL-KDD dataset

*C. Machine Learning Algorithms*

For evaluation purpose, this project has considered Bayes Network, Decision Table, Decision tree, J48, K-Nearest Neighbour, Random Forest and Random Tree algorithms.

*1) Bayesian Network*: is an efficient probabilistic graphical model for event classification scheme [18]. After some development in this are Jemili et. al. [19] proposed that Bayesian networks can be used to build a framework for an adaptive IDSs.

*2) Decision Table*: is one of the simplest hypothesis spaces possible for supervised learning algorithms [20]. Later Chen et. al. [21] proposed a hybrid classifier that combined decision table and Naïve Bayes, which selected the deterministic attributes and outperformed in comparison to both the techniques considered separately.

*3) Decision Tree:* is a tree-like structure that has it's each leaf representing a classification category. A decision tree with a range of discrete (symbolic) class labels is called a classification tree, whereas a decision tree with a range of continuous (numeric) values is called a regression tree. Many improvements were made in the performance of decision tree and eventually, a multiclass intrusion detection system was proposed using decision tree [22].

*4) J48:* is an open source Java version of the C4.5 algorithm [23]. The output of this classifier is in the form of decision binary trees but with more stability between computation time and accuracy.

*5) K-nearest neighbor:* is lazy learning algorithm and traditional non-parametric technique to classify a point in the sample based on the k nearest neighbors of that data point. Many improvements and hybrid classifiers have been proposed till date such as Lin et. al. [24] proposed a hybrid of cluster center and nearest neighbor showing a better performance in terms of accuracy, detection rates, and false alarms.

*6) Random Forest:* is a machine learning method that combines the decision trees and ensemble learning [25]. Then the prediction is decided by majority voting. Later Zhang et.al. [7] proposed a hybrid system that combines the advantages of the misuse and anomaly detection mechanism of random forest in IDS, which improved the overall performance of IDS.

*D. Evaluation Metrics*

For the performance analysis of all the algorithms considered in the project, we recorded true positive, false positive, precision, recall and F-measure.

*1) True Positive:* is an outcome where the model correctly predicts the positive class.

*2) False positive:* is a result that indicates that a given condition is present when it is not.

*3) Precision:* is the ability of a classification model to identify only the relevant data points.

$$\text{Precision} = \frac{tp}{tp + fp}$$

*4) Recall:* is the ability of a model to find all the relevant cases within a dataset.

$$\text{Recall} = \frac{tp}{tp + fn}$$

*5) F-measure*: is a measure of test's accuracy and is defined as the weighted mean of the precision and recall of the test.

$$\text{Accuracy} = \frac{tp + tn}{tp + tn + fp + fn}$$

IV. PROPOSED HYBRID MODEL

We proposed a new hybrid model for network intrusion detection, which aims to achieve better precision, recall and accuracy (F-measure). To reify this, we combined Decision Tree and Random Forest algorithms using stacking scheme.

Stacking scheme is an ensemble learning technique which combines multiple classification models with the help of meta-classifier. The stacking process involves two phases: first, training of each selected model on complete dataset provided initially as input; followed by the training of meta-level classifier to fit it based on the outputs (meta-features) of the individual classification models in the ensemble obtained from the previous step (Fig. 4).

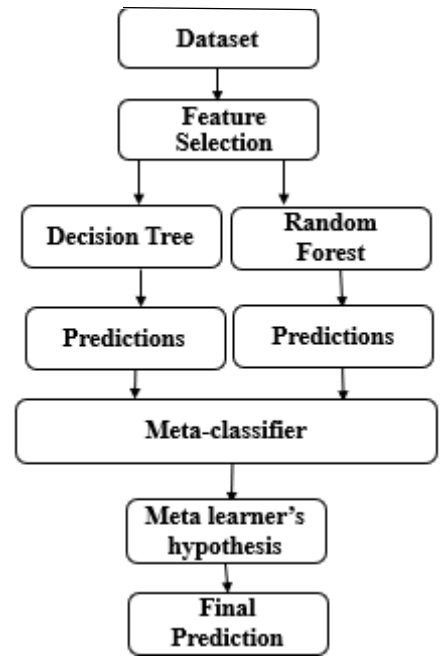

**Fig.4.** Stacking Scheme Architecture

The main benefit of using stacking scheme to make a hybrid model is that, if one of the base classifiers has been incorrectly learned in the first phase then the second phase (meta-classifier) might be able to detect this and apply correct training.

The proposed model's pseudo code can be viewed as below:

| |
|---|
| *Algorithm 1*: Proposed Model |
| 1:    Procedure model () |
| 2:    Input = Dataset |
| 3:    Apply Information Gain Attribute Evaluation method to reduce the number of features such that the information gain |

| | |
|---|---|
| 4: | Use stacking scheme for combining algorithms (Decision Tree and Random Forest) for phase 1 |
| 5: | Define meta-classifiers using stacking scheme for phase 2 and implement Decision Tree and Random Forest for this |
| 6: | Again, define meta-classifier for the stacking scheme in phase 2 by using Decision Tree |
| 7: | Propose the hybrid model |
| 8: | Provide the dataset to this hybrid model |
| 9: | Class with highest probability is predicted by the model |

## V. RESULT ANALYSIS

The experiments were conducted on Intel$^R$ Core$^{TM}$ i5-2450m CPU, 2.50ghz 2.50 GHz and 4GB RAM machine. Initially the experiments were performed to evaluate the performance of few machine learning algorithms used for intrusion detection systems, namely, Bayes Network, Decision Table, Decision tree, J48, K-Nearest Neighbour, Random Forest and Random Tree.

When the above-mentioned algorithms are tested with CICIDS2017 dataset, values of true positive, false positive, precision, recall and F-measure are recorded as results (TABLE IV). We can observe from TABLE IV that Bayes Network performed relatively poor in comparison to other algorithms, having highest false positive (0.075) and lowest precision (0.976) and accuracy (0.976). In contrast, K-Nearest Neighbor, Random Forest and Random Tree showed performance with having the highest true positive (0.98), precision (0.98) and accuracy (0.98). Even though the overall performance of all these algorithms are very high (close to 98%), but tasks involving critical information may not be able to afford even the slightest intrusion.

Following the above experiments, next the experiments were conducted using NSL-KDD dataset. The results obtained are illustrated using TABLE V. According to TABLE V, Decision Table's performance was low in comparison to the other algorithms followed by Bayes Network. Precision, recall and F-measure for Decision Table were recorded as 0.814, 0.726 and 0.718 respectively. Whereas, Decision Tree outperformed all the other algorithms, with precision, recall and F-measure recorded as 0.848, 0.833 and 0.833 respectively.

Considering our hybrid detection model, when it was tested using NSL-KDD dataset, it performed better than the other algorithms considered for analysis in this project. It achieved the highest F-measure (0.852), true positive (0.852), recall (0.852) and precision (0.862), and the lowest false positive rate (0.134) (TABLE VI) amongst all the classifiers. Thus, it shows that the new model can be used in IDSs, delivering better results.

When the hybrid detection model is tested on CICIDS2017, it was noticed that its performance was similar to K-Nearest Neighbor, Random Forest and Random Tree (TABLE VI). Thus, showing that our model was able to achieve the best results similar to the best algorithms for CICIDS2017 dataset and the best overall results in comparison to other algorithms for NSL-KDD dataset.

TABLE IV: PERFORMANCE COMPARISON FOR CICIDS2017

| Algorithms | TP | FP | Precision | Recall | F-measure |
|---|---|---|---|---|---|
| Bayes Network | 0.976 | 0.075 | 0.976 | 0.976 | 0.976 |
| Decision Table | 0.980 | 0.071 | 0.980 | 0.980 | 0.980 |
| Decision tree | 0.980 | 0.070 | 0.980 | 0.980 | 0.979 |
| J48 | 0.980 | 0.071 | 0.980 | 0.980 | 0.979 |
| K-Nearest Neighbor | 0.980 | 0.069 | 0.980 | 0.980 | 0.980 |
| Random Forest | 0.980 | 0.069 | 0.980 | 0.980 | 0.980 |
| Random Tree | 0.980 | 0.069 | 0.980 | 0.980 | 0.980 |

TABLE V: PERFORMANCE COMPARISON FOR NSL-KDD

| Algorithms | TP | FP | Precision | Recall | F-measure |
|---|---|---|---|---|---|
| Bayes Network | 0.742 | 0.203 | 0.819 | 0.742 | 0.736 |
| Decision Table | 0.726 | 0.214 | 0.814 | 0.726 | 0.718 |
| Decision Tree | 0.833 | 0.148 | 0.848 | 0.833 | 0.833 |
| J48 | 0.783 | 0.172 | 0.838 | 0.783 | 0.781 |
| K-Nearest Neighbour | 0.805 | 0.155 | 0.851 | 0.805 | 0.803 |
| Random Forest | 0.802 | 0.158 | 0.847 | 0.802 | 0.801 |
| Random Tree | 0.815 | 0.160 | 0.838 | 0.815 | 0.815 |

TABLE VI: PERFORMANCE COMPARISON FOR HYBRID MODEL

| Hybrid Model | TP | FP | Precision | Recall | F-measure |
|---|---|---|---|---|---|
| CICIDS2017 | 0.98 | 0.069 | 0.980 | 0.980 | 0.980 |
| NSL-KDD | 0.852 | 0.134 | 0.862 | 0.852 | 0.852 |

## VI. CONCLUSION

Results of the analysis of algorithms using CICIDS2017 illustrated that this data can be used to test and train the intrusion detection systems since, CICIDS2017 have records that are actual network traffic and it helps in better training of the classifiers. Hence, resulting in better detection accuracy, precision and recall.

The proposed hybrid model performs similar to the well-known algorithms with high accuracy, recall and precision when tested against CICIDS2017 dataset. Since this dataset provides well representation of real network traffic, hence this shows that our model can be used for practical implementation of IDSs. When it was tested using NSL-KDD dataset, our hybrid model gave better overall results. Thus, ensuring that our new model can be used for implementing intrusion detection systems.

REFERENCES

[1] I. Sharafaldin, A. H. Lashkari and A. A. Ghorbani, "Toward Generating a New Intrusion Detection Dataset and IntrusionTraffic Characterization," in *4th International Conference on Information Systems Security and Privacy (ICISSP)*, 2018.

[2] M. Ahmed, A. N. Mahmood and J. Hu, "A survey of network anomaly detection techniques," *Journal of Network and Computer Applications,* vol. 60, pp. 19-31, 2016.

[3] D. E. Denning, " An intruison-detection model," in *IEEE Symposium on Security and Privacy*, 1987.

[4] J. McHugh, "Testing Intrusion detection systems: a critique of the 1998 and 1999 DARPA intrusion detection system evaluations as performed by Lincoln Laboratory," in *ACM Transactions on Information and System Security*, 2000.

[5] Y. Liao and V. R. Vemuri, "Use of K-nearest neighbor classifier for intrusion detection," *Computer and Security,* vol. 21, no. 5, pp. 439-448, 2002.

[6] M. Mahoney and P. Chan, "An analysis of the 1999 DARPA / Lincoln laboratory evaluation data for network anomaly detection," in *Recent Advances in Intrusion Detection, 6th International Symposium, RAID 2003, Pittsburgh, PA, USA*, 2003.

[7] J. Zhang, M. Zulkernine and A. Haque, "Random-forests-based network intrusion detection systems," *IEEE Transactions on Systems, Man, and Cybernetics, Part C (Applications and Reviews),* vol. 38, no. 5, pp. 649 - 659, 2008.

[8] U. Ravale, N. Marathe and P. Padiya, "Feature selection based hybrid anomaly intrusion detection system using K means and RBF kernel function," in *Proceeding of International Conference on Advanced Computing* , 2015.

[9] S. Aljawarneh, M. Aldwairi and M. B. Yassein, "Anomaly-based intrusion detection system through feature selection analysis and building hybrid efficient model," *Journal of Computational Science,* vol. 25, pp. 152-160, 2018.

[10] M. C. Belavagi and B. Muniyal, "Performance Evaluation of Supervised Machine Learning Algorithms for Intrusion Detection," *Procedia Computer Science,* vol. 89, pp. 117-123, 2016.

[11] M. Almseidin, M. Alzubi, S. Kovacs and M. Alkasassbeh, "Evaluation of Machine Learning Algorithms for Intrusion Detection System," in *15th International Symposium on Intelligent Systems and Informatics*, 2017.

[12] M. Zaman and C. H. Lung, "Evaluation of Machine Learning Techniques for Network Intrusion Detection," in *IEEE/IFIP Network Operations and Management Symposium*, 2018.

[13] "Weka 3: Data Mining Software in Java," [Online]. Available: https://www.cs.waikato.ac.nz/ml/weka/. [Accessed 08 04 2019].

[14] "Intrusion Detection Evaluation Dataset (CICIDS2017)," [Online]. Available: https://www.unb.ca/cic/datasets/ids-2017.html. [Accessed 08 04 2019].

[15] "NSL-KDD dataset," [Online]. Available: https://www.unb.ca/cic/datasets/nsl.html. [Accessed 08 04 2019].

[16] M. Tavallaee, E. Bagheri, W. Lu and A. Ghorbani, "A Detailed Analysis of the KDD CUP 99 Data Set," in *Second IEEE Symposium on Computational Intelligence for Security and Defense Applications (CISDA)*, 2009.

[17] S. Revathi and A. Malathi, "A Detailed Analysis on NSL-KDD Dataset Using Various Machine Learning Techniques for Intrusion Detection," *International Journal of Engineering Research & Technology,* vol. 2, no. 12, 2013.

[18] C. Kruegel, D. Mutz, W. Robertson and F. Valeur, "Bayesian event classification for intrusion detection," in *Proceedings of 19th annual computer security applications conference*, 2003.

[19] F. Jemili, M. Zaghdoud and A. Ben, "A framework for an adaptive intrusion detection system using Bayesian network," in *IEEE Intelligence and Security Informatics*, 2007.

[20] R. Kohavi, "The power of decision tables," in *ECML'95 Proceedings of the 8th European Conference on Machine Learning*, 1995.

[21] C. Chen, G. Zhang, J. Yang, J. Milton and A. D. Alcántarad, "An explanatory analysis of driver injury severity in rear-end crashes using a decision table/Naïve Bayes (DTNB) hybrid classifier," *Accident Analysis & Prevention,* vol. 90, pp. 95-107, 2016.

[22] S. A. Mulay, P. R. Devale and G. V. Garje, "Intrusion Detection System using Support Vector Machine and Decision Tree," *International Journal of Computer Applications,* vol. 3, no. 3, 2010.

[23] R. Quinlan, C4.5: Programs for Machine Learning, San Mateo, CA, USA: Morgan Kaufmann, 1993.

[24] W. C. Lin, S. W. Ke and C. F. Tsai, "CANN: An intrusion detection system based on combining cluster centers and nearest neighbors," in *Knowledge-based systems*, 2015.

[25] L. Breiman, "Random Forests," *Machine Learning,* vol. 45, no. 1, pp. 5-32, 2001.

[26] J. Zhang, M. Zulkernine and A. Haque, "Random-forests-based network intrusion detection systems," in *IEEE TRANSACTIONS ON SYSTEMS, MAN, AND CYBERNETICS*, 2008.